\documentstyle[12pt]{article}
\newcommand{\draft}{%           Normal spacing
        \renewcommand{\baselinestretch}{1.0}%
        \small\normalsize%
}

\draft
\begin{document}
\title{\bf Influences of neutron star parameters on evolutions of 
different types of pulsar; evolutions of anomalous X-ray pulsars, soft 
gamma repeaters and dim isolated thermal neutron stars on the P-\.{P} 
diagram} 
\author{Oktay H. Guseinov$\sp{1,2}$
\thanks{e-mail:huseyin@gursey.gov.tr},
A\c{s}k\i n Ankay$\sp1$
\thanks{e-mail:askin@gursey.gov.tr},
Sevin\c{c} O. Tagieva$\sp3$
\thanks{email:physic@lan.ab.az}, \\ \\
{$\sp1$T\"{U}B\.{I}TAK Feza G\"{u}rsey Institute} \\
{81220 \c{C}engelk\"{o}y, \.{I}stanbul, Turkey} \\
{$\sp2$Akdeniz University, Department of Physics,} \\
{Antalya, Turkey} \\
{$\sp3$Academy of Science, Physics Institute, Baku 370143,} \\
{Azerbaijan Republic}}
%\\ \\ \\ \\ \\ \\
%{3 copies of: ?? pages of text + 4 pages of Tables.} \\ \\
%{Please send proofs to:} \\
%{A\c{s}k\i n Ankay} \\
%{Orta Do\u{g}u Teknik \"{U}niversitesi} \\
%{Fizik B\"{o}l\"{u}m\"{u}} \\
%{06531-Ankara} \\
%{T\"{u}rkiye}}

\date{}
\maketitle
%\final
\begin{abstract}
\noindent
Influences of the mass, moment of inertia, rotation, absence of stability 
in the atmosphere and some other parameters of neutron stars on the 
evolution of pulsars are examined. It is shown that the locations and 
evolutions of soft gamma repeaters, anomalous X-ray pulsars and other 
types of pulsar on the period versus period derivative diagram can be 
explained adopting values of B$<$10$^{14}$ G for these objects. This 
approach gives the possibility to explain many properties of different 
types of pulsar.   
\end{abstract}

\section{Introduction}
Soft gamma repeaters (SGRs) and anomalous X-ray pulsars (AXPs) have been 
observed in the last about 35 years, but their important role in 
astrophysics has been understood only about 10 years ago and intensive 
observations and theoretical research have since been done on these exotic 
objects. Later, some other types of neutron star have been identified. 
These are dim radio quiet neutron stars (DRQNSs), which are connected to 
supernova remnants (SNRs), and the single neutron stars with very long 
spin periods which are named as dim isolated thermal neutron stars 
(DITNSs). It is possible to find many data on these sources given in 
Mereghetti \& Stella (1995), Mereghetti et al. (2002), Becker \& 
Aschenbach (2002), and Guseinov et al. (2003a, 2003b).

The existence of single neutron stars with large values of P$>$5 s and 
\.{P}$>$5$\times$10$^{-13}$ s/s was not expected in pulsar astronomy. In 
addition to this, the inverse value of the ratio of the X-ray luminosity 
(L$_x$) to the rate of rotational energy loss (L$_x$/\.{E}$>$10) for AXPs 
and SGRs and the $\gamma$-ray bursts observed from SGRs were unexpected 
phenomena (Thompson \& Duncan 1995; Mereghetti 2001). Today, it is 
natural to put AXPs and SGRs together as one type of young neutron star 
as done in the magnetar model. 

It is easy and a usually chosen way to explain $\gamma$-ray 
bursts by recombination of the magnetic field and this is indeed  
a more reliable way. On the other hand, the initial magnetic field 
B$\sim$10$^{14}$-10$^{15}$ G may easily explain the ages of AXPs and SGRs 
and also the burst energy and the time characteristics of the bursts
together with the temperatures and luminosities of these objects 
(Thompson \& Duncan 1995; Mereghetti et al. 2002). This approach was also 
important to expand the limit of the physical idea and the application 
used in astrophysics.

The processes of steady accretion onto neutron stars have begun to be 
examined since $\sim$40 years ago and the idea of accretion was 
successfully used to predict binary X-ray sources together with their 
main properties (Zeldovich \& Guseinov 1966; Shklovskii 1967; Guseinov 
1970; Bisnovatyi-Kogan \& Komberg 1974; Lipunov 1992). Naturally, the 
theory of accretion from fall-back matter (Chatterjee et al. 2000) was 
used together with neutron star activity to explain the nature of 
AXPs/SGRs (Rothschild et al. 2002a, 2002b, but see also Tagieva et al. 
2003). 

In this work, we analyse possible evolutions of different types of single 
pulsar on the P-\.{P} diagram in detail taking into consideration the 
possibility of the absence of stability for different masses of neutron 
stars. We consider AXPs/SGRs and DITNSs to be neutron stars with initial 
magnetic field $\ge$10$^{13}$ G. 

\section{Influences of the value of neutron star mass on electrodynamical 
parameters of pulsars}
There are relations between mass and radius of neutron star which are 
constructed using models with different equations of state. 
Maximum mass values of the theoretical mass-radius relations lie in the 
interval 1.5-2.2 M$_{\odot}$ for which the radius value changes in the 
interval 10-12 km (Bejger \& Haensel 2003). 

Moment of inertia (I) of a star is equal to $\alpha$MR$^2$, where $\alpha$ 
is a coefficient which depends on the mass density (the total energy  
density) with respect to radius, M the mass and R the radius of the star. 
From the relations given in Bejger \& Haensel (2003) and 
Ravenhall \& Pethick (1994) it is seen that the coefficient 
changes monotonically with respect to mass and adopting a radius of 11 km 
it takes values of 0.37 and 0.44 (corresponding to 1.5 M$_{\odot}$ and 2.2 
M$_{\odot}$ respectively) for neutron stars. As seen, the value of 
$\alpha$ weakly depends on the equation of state and on the mass of the 
compact star. The value of $\alpha$ slowly decreases with the diminishing 
of the compactness. Using these values we have estimated: 
I$\cong$1.3$\times$10$^{45}$ gr cm$^2$ for $\alpha$=0.37 and M=1.5 
M$_{\odot}$. 

The neutron star mass values which are estimated from binary systems 
have a sharp 
peak at 1.4 M$_{\odot}$ (see e.g. Tagieva et al. 2000; Lyne \& 
Graham-Smith 1998; Charles \& Coe 2003). It is not possible to measure 
the mass of single neutron stars, but one may assume that single pulsars 
also have masses about 1.4 M$_{\odot}$. Please note that, on the period 
versus period derivative (P-\.{P}) diagram of pulsars, the lines of 
constant effective magnetic field (B) and constant 
characteristic time ($\tau$), which are found using P and \.{P} values, 
are shown assuming R=10 km and I=10$^{45}$ gr cm$^2$ (see e.g. Lyne \& 
Graham-Smith 1998) for the case of pulsars in vacuum which have pure 
magnetic dipole radiation (n=3, where n is the braking index). 
The lines of constant rate of rotational energy loss (\.{E}) are also 
shown on the P-\.{P} diagram (Figure 1). 
Even if there exist considerable differences in the masses of neutron 
stars, the influences of these differences on B, $\tau$ and \.{E} are 
neglected. Actually, in some cases the real B value must be less than the 
effective B value found by assuming the case of pure magnetic dipole 
radiation. Likewise, the characteristic time must be different than the 
real age in some cases and moreover the values of $\tau$ may be different 
for two pulsars with the same values of P and \.{P}. 
On the other hand, in contrast to the cases of B and $\tau$, the \.{E} 
values found from P and \.{P} values are always the actual values of 
pulsars if we neglect the small differences in the values of I for 
different pulsars. This is true also because of the very fast change in 
\.{E} along the evolutionary tracks of pulsars.

The maximum values of the binding energy (defect mass) 
correspond to the maxima of the mass versus central density relations.
In these relations, the part on the right of the maximum (i.e. large 
central density part) corresponds to the unstable states of neutron star, 
whereas, the part on the left of the maximum corresponds to the stable 
states of neutron star and the most stable neutron star states are at and 
just on the left side of the maximum (Zeldovich \& Novikov 1971). 
The value of 1.4 M$_{\odot}$ found from observations is close to the 
lowest value (1.5 M$_{\odot}$) of the interval of the maximum mass given 
above. So, the neutron star models with maximum mass values $\sim$1.5-1.6
M$_{\odot}$ are reliable, whereas, the probability of the existence of 
neutron stars with masses $\sim$2-2.2 M$_{\odot}$ is very low. 
In this case, the I value corresponding to the maximum mass value must 
be adopted as $\sim$(1.3-1.4)$\times$10$^{45}$ gr cm$^2$ and this is 
very close to the commonly adopted value of 10$^{45}$ gr cm$^2$. 
It is necessary to note that even rapidly rotating neutron stars with 
various different equations of state based on general theory of 
relativity may have additional masses only up to about 10\%. Influence on 
the radius and especially on the equatorial region is slightly larger in 
such cases (Gurovich \& Guseinov 1965).  

Let us consider a 0.7-0.8 M$_{\odot}$ neutron star and a 1.5 M$_{\odot}$ 
neutron star both of 
which have the same value of the component of dipole magnetic field B 
perpendicular to the rotation axis and both of which are located in 
vacuum. Magnetic dipole radiation of neutron star (Lipunov 1992) 
\begin{equation}
L = \frac{2}{3} \frac{\mu^2 \omega^4}{c^3}Sin^2\beta = \frac{1}{3} 
\frac{B^2 R^6 \omega^4}{c^3}Sin^2\beta
\end{equation} 
(where $\mu$ is the magnetic moment, $\omega$ the angular velocity, c the 
speed of light, and $\beta$ the angle between the rotation axis and the 
axis of the magnetic field) leads to braking of the pulsar spin. If we 
equate the magnetic dipole luminosity (Eqn.1) to the rate of rotational 
energy loss of neutron stars which have rigid body rotation given as
\begin{equation}
\dot{E}=\frac{4\pi ^2 I \dot{P}}{P^3}
\end{equation}
then we get
\begin{equation}
\dot{P} \propto \frac{B^2 R^4}{M P}.
\end{equation} 
As seen from the equation, for the same values of B and P, the \.{P} value 
of a 0.7-0.8 M$_{\odot}$ neutron star is about 3-4 times larger than the 
\.{P} value of a 1.5 M$_{\odot}$ neutron star considering also the 
possible increase in 
the radius value up to a factor of $\sim$1.2 when we reduce the mass 
value. So, the \.{E} value of a 1.5 M$_{\odot}$ neutron star must roughly 
be a factor of $\sim$2.5 less than the \.{E} value of a 0.7-0.8 
M$_{\odot}$ neutron star for a given period value.
 
If we use the observational values of P and \.{P}, the component of the 
effective magnetic field perpendicular to the spin-axis is determined as 
\begin{equation}
B=\sqrt{\frac{3c^3IP\dot{P}}{8\pi ^2 R^6}}.
\end{equation}
This value of B is equal to the real value if all the other mechanisms for 
the rotational energy loss are negligible compared to the magnetic dipole 
radiation. 

The characteristic time of a neutron star which is defined as
\begin{equation}
\tau = \frac{P}{(n-1)\dot{P}}
\end{equation}
(where n is the braking index) does not depend on the parameters of 
neutron star, which may change in time, nor on any conditions 
in the environment if n is assumed to be constant. In the case of 
pulsars which rotate as rigid body and which are in hydrodynamical 
equilibrium in vacuum with pure magnetic 
dipole radiation (n=3), if the initial spin period (P$_0$) is negligible 
compared to P, then $\tau$ is equal to the real age of the pulsar. 
If n=constant along all the evolutionary tracks, then the age of the 
pulsar is
\begin{equation}
t = \frac{P}{(n-1)\dot{P}} (1-(\frac{P_0}{P})^{n-1}).
\end{equation}
In principle, the value of n may change continuously along the 
evolutionary track. As $\tau$ shows the character of the period 
change at every point along the trajectory, it is a local characteristic 
and expression (5) always gives the real value of $\tau$ if we use the 
real value of n at each point of the evolutionary track for each pulsar. 

There will be a difference of a factor of about 3-4 between the $\tau$ 
values of a 1.5 M$_{\odot}$ neutron star and a 0.7-0.8 M$_{\odot}$ neutron 
star because of the larger value of \.{P} for smaller value of mass for 
the neutron star if both of the neutron stars have the same magnetic 
field. 

Let us consider two neutron stars which have the same dipole magnetic 
field but different masses under the following assumption. Both of these 
rigidly rotating neutron stars have the same rotational period, but the 
less massive one with mass M$_2$ has  
10 times larger \.{P} compared to the other. In such a case, 
$\tau$$_2$=0.1$\tau$$_1$ from expression (5). If we estimate value of 
the ratio E$_2$/E$_1$ from Figure 1, we find that E$_2$=10E$_1$. 
Actually, if \.{P}$_2$=10\.{P}$_1$ then E$_2$$>$E$_1$ but not 10 times, 
because the ratio E$_2$/E$_1$ depends on the equation of state which is 
used in the constructed model of neutron star. If the neutron star does 
not rotate as rigid body, then the dependence on the total value of 
rotational moment becomes stronger. 

In Figure 1, the locations of $\tau$=constant lines do not change if 
n=constant as seen from equation (5), but the 
locations of and the distances between the lines of B=constant and 
\.{E}=constant must change as seen from the dependences (2) and (4). The 
change in mass values and the relation between the mass and the radius 
lead to a change in the value of I and a change in the relation between 
$\tau$ and the real age. It is well known that for the neutron stars which 
are in hydrodynamical equilibrium, both the coefficient in the expression 
I=$\alpha$MR$^2$ and the value of R depend very weakly on the equation of 
state. But these quantities may considerably be different when the neutron 
star is far from equilibrium, has a large rotational moment and magnetic 
field strength, and/or does not rotate as rigid body.     

Electric field intensity which is generated under the rotation of a 
neutron star which has pure magnetic dipole radiation and around which no 
plasma exists may roughly be given as (Lipunov 1992)
\begin{equation}
E_{el} \sim \frac{BR}{P}.
\end{equation}
Therefore, for equal values of P and B, the star with a larger radius 
(smaller mass) must have a larger value of E$_{el}$. The exact expression 
for E$_{el}$ may be found only in simple model approximations. But it is 
enough to know the simple relation (7) for our purposes. 

Using equation (4) we get
\begin{equation}
E_{el} \sim \frac{I^\frac{1}{2}}{R^2} (\frac{\dot{P}}{P})^\frac{1}{2}.
\end{equation}
Therefore, the E$_{el}$ value 
of a 1.5 M$_{\odot}$ neutron star must be a factor of about 1.1-1.2 
less than the E$_{el}$ value of a 0.7-0.8 M$_{\odot}$ neutron star. 
It must be noted that, in actuality, there can be some plasma around the 
neutron star. There may be jets also. These can distort the magnetic field 
so that the expression for E$_{el}$ can be different. 
 
From equations (5) and (8) we get
\begin{equation}
E_{el} \sim I^\frac{1}{2} R^{-2} \tau ^{-\frac{1}{2}}
\end{equation}
so that, using logarithmic scale, the average constant E$_{el}$ lines will 
be parallel to constant $\tau$ lines on the P-\.{P} diagram for fixed 
values of I and R. 

\section{Which changes may take place on the P-\.{P} diagram under the 
change of neutron star parameters?}
The lines of constant B, $\tau$ and \.{E} pass through with given 
parameters as displayed in Figure 1 and with fixed intervals next to each 
other in the ideal case. The 'ideal case' means that the neutron stars 
are assumed to have $\sim$1 M$_{\odot}$ (under the chosen equation of 
state), having pure dipole magnetic field with constant values, rotating 
as rigid body, and being 
in hydrodynamical equilibrium. Actually, we can not expect that all  
neutron stars have the same mass and radius, have similar configurations 
for the magnetic field, locate in similar environments nor they are in 
hydrodynamical equilibrium. Some of them can reach the hydrodynamical 
equilibrium state in short time and some in relatively long time. 
The conditions mentioned above can be  
different especially for young neutron stars and in this case we can 
expect them to have very different evolutionary tracks on the P-\.{P} 
diagram. 

We can estimate only a value of \.{E} close to the exact value from the 
position of the pulsar on the P-\.{P} diagram. Two pulsars which are 
located very close to each other on the P-\.{P} diagram may have 
different values of $\tau$, B, E$_{el}$ and age as seen from expressions 
(5), (4), (8) and (6). 

In general, the neutron star model with R=10 km and I=10$^{45}$ gr cm$^2$ 
is used and therefore the lines of B=constant and \.{E}=constant on the 
P-\.{P} diagram are constructed by using equations
\begin{equation}
B = (\frac{3c^3 I}{8\pi^2 R^6})^\frac{1}{2} (P\dot{P})^\frac{1}{2} = 
3.2\times 10^{19} (P\dot{P})^\frac{1}{2} \hspace{0.2cm} G
\end{equation}
and
\begin{equation}
\dot{E} = 4\pi^2 I \frac{\dot{P}}{P^3} = 3.94\times 
10^{46}\frac{\dot{P}}{P^3} \hspace{0.2cm} erg/s
\end{equation}
(Lyne \& Graham-Smith 1998). For neutron stars with masses about 0.7 
M$_{\odot}$, it is possible to use more reliable values of R=12 km and 
I=7.2$\times$10$^{44}$ gr cm$^2$ instead of the very simple values of R 
and I, though in principle there will be no significant change. In this 
case, instead of the expressions (10) and (11), we have 
\begin{equation}
B = 1.58\times 10^{19} (P\dot{P})^\frac{1}{2} \hspace{0.2cm} G 
\end{equation}
and
\begin{equation}
\dot{E} = 2.84\times 10^{46}\frac{\dot{P}}{P^3} \hspace{0.2cm} erg/s
\end{equation}
similar to the expressions (10) and (11) which are always used for 
the very compact and stable neutron stars. The relations (10) and (11) are 
also applicable for dominant number of neutron stars which have masses 
1-1.5 M$_{\odot}$. If we use the expressions (12) and (13) for the 
P-\.{P} diagram, the locations of $\tau$=constant lines do not change, the 
lines of \.{E}=constant practically do not change either, but the lines 
of B=constant shift down by a factor of 2. Therefore, a more accurate 
definition of the places of these lines on the P-\.{P} diagram does not 
have an essential role and can not help us in solving the important 
problems about the origins and evolutions of different types of neutron 
star. 

As seen from Figure 1, bulk of the neutron stars are born with magnetic 
field similar to the effective field of Crab and Vela pulsars 
(10$^{12}$-10$^{13}$ G). We can assume that 
about 20\% of neutron stars are born with masses about half of the 1.4 
M$_{\odot}$ which is the maximum value in the number-mass distribution of 
neutron stars in binary systems (Lyne \& Graham-Smith 1998; Tagieva et al. 
2000; Charles \& Coe 2003). These neutron stars may have R=13 km as found 
from the soft equations of state of rotating neutron stars (see Bejger \& 
Haensel 2003). In this case, \.{P} value of 
such a pulsar must be about 5-6 times larger than the \.{P} of 
a normal radio pulsar (similar to Crab) if the dipole magnetic field and 
the P values of these pulsars are the same (see equation (3)). The 
small mass neutron stars with such values of \.{P} must also have about 
5-6 times smaller value of $\tau$ compared to the bulk group of pulsars 
with the same value of P. As seen from Figure 1, there will be no problem 
in explaining the position of DITNS RXJ0720.4-3125 on the P-\.{P} diagram 
by using such an approach (i.e. considering such objects as neutron stars 
with smaller masses). This approach can be useful in explaining most of 
the DITNSs. Most of these sources may have values of P$<$10 s and 
\.{P}$<$10$^{-12}$ s/s. 

\section{Some problems about AXPs/SGRs and DITNSs}
Both the magnetar model and the accretion from fall-back matter model 
have some considerable successes in explaining some properties of 
AXPs/SGRs as mentioned briefly in the Introduction. We suggest to bring 
the successful parts of both approaches together and in addition to this 
we propose a way to explain the properties of these objects 
based on the intrinsic parameters (different mass, different structure, 
absence of hydrodynamical equilibrium etc.). Guseinov et al. 
(2003a, 2003b) analyse the data of AXPs/SGRs and suggest the initial 
magnetic field to be 3$\times$10$^{13}$-10$^{14}$ G by examining also the 
activity of young neutron stars. From the analysis 
of the X-ray properties and the birth rates of different types of neutron 
star (Guseinov et al. 2003c), a possibility arises to step back from the 
idea of very high magnetic field. 

In Figure 1, all the 9 radio pulsars with distances up to 3.5 kpc from the 
Sun which are connected to SNRs and have values of 
$\tau$$\le$5$\times$10$^4$ yr are represented. Among these 9 pulsars, SNR 
shell and pulsar wind nebula (PWN) have not been observed for 4 and 3 of 
them, respectively (Green 2001; Guseinov et al. 2003d). The 
ages of SNRs in many cases are in agreement with the $\tau$ values of the 
pulsars connected to them in the error limits and their ages in some cases 
may practically be as large as 3$\times$10$^4$ yr with the exception of 
the pair pulsar J0205+6449 -- SNR G130.7+3.1. The $\tau$ value of this 
pulsar is 5.4$\times$10$^3$ yr, but the SNR connected to this pulsar may 
be a historical remnant with an age of $\sim$800 yr (Lorimer et 
al. 1998). Since P=0.066 s for this pulsar and as the remnant is 
F-type (i.e. a reliable age estimation can not be made for the SNR), the 
existing discrepancy can be removed. Therefore, it is not necessary to 
discuss the differences between the values of $\tau$ and the real ages 
for this group of pulsars when discussing the average characteristics. 
However, more reliable larger values of age can be adopted based on the 
data of the SNRs which are connected to this group of radio pulsars:
t$\sim$3$\times$10$^4$ yr (see Guseinov et al. 2003e and the 
references therein). SNR G8.7-0.1 has such a large age of about 
3$\times$10$^4$ yr (Finley \& Ogelman 1994) and this SNR is connected to 
pulsar J1803-2137 (Frail et al. 1994). In general, the positions of these 
9 radio pulsars on the P-\.{P} diagram do not raise any questions. All of 
them are located in the belt B=(2.5-10)$\times$10$^{12}$ G. 

There are 7 SNRs with ages up to 2$\times$10$^4$ yr which contain DRQNSs 
in the same region of 3.5 kpc from the Sun (Guseinov et al. 2003c, 2003d). 
The locations of 2 of them, 1E1207.4-5209 and RXJ0002+6246, on the 
P-\.{P} diagram are known with small errors (see Figure 1). For all the 
others, even the value of P is not known that we can not find the 
positions of them on the P-\.{P} diagram. On the other hand, we know that 
PWN exists only around the pulsars which have 
\.{E}$>$3$\times$10$^{35}$ erg/s and L$_{2-10keV}$$>$10$^{33}$ erg/s 
(Guseinov et al. 2003d). Since DRQNSs do not have such large values of 
L$_{2-10keV}$ and no PWN has been found around them, they must be located 
on the right hand side of  
\.{E}=3$\times$10$^{35}$ erg/s line in Figure 1. Therefore, the values of 
$\tau$ for DRQNSs, on average, can be about one order (maybe more) of 
magnitude larger than their real ages. It follows from the numbers and 
ages of the SNRs which are connected to radio pulsars and DRQNSs that the 
birth rates of these objects are approximately the same.
According to Guseinov et al. (2003c), the birth rate of DRQNSs 
near the Sun is about half of the birth rate of all pulsars. These DRQNSs 
may directly be born at their present locations on the P-\.{P}      
diagram or they may come to these positions in very short time after the
birth. There is no data directly showing if pulsars can be born with
initial periods greater than 0.1 s or not.

On the P-\.{P} diagram (Figure 1) in the region B=(4-11)$\times$10$^{11}$ 
G and $\tau$=(1-6)$\times$10$^5$ yr, there are also 5 radio pulsars from 
which cooling X-ray radiation have been detected. The birth rate of these 
pulsars is 0.06 of the birth rate of all radio pulsars. This is about 0.04 
of the combined birth rate of all the radio pulsars and the DRQNSs  
together (Guseinov et al. 2003c). Two of these sources, J1952+3252 and 
J0538+2817, are connected to S-type SNRs. We must add to this group also 
the radio pulsar J0659+1414 which is connected to a SNR and maybe the 
Geminga pulsar (see their positions on the P-\.{P} diagram). In this 
case, the birth rate of pulsars with $\tau$$>$10$^5$ yr and P$<$0.4 s 
from which cooling X-ray radiation have been observed will be 0.084 
of the birth rate of all pulsars. As seen, this value is very small. On 
the other hand, these pulsars avoid the belt B=10$^{12}$-10$^{13}$ G. 
These data also show that it is necessary to give up the commonly 
used simple approach in order to be able to explain the pulsar evolution 
on the P-\.{P} diagram. Naturally, future observations must lead to an 
increase in the number density of cooling pulsars. Note that, the ages 
are assumed to be equal to the values of $\tau$ for these 5 pulsars in 
the cooling theories (Kaminker et al. 2002; Yakovlev et al. 2002). 

Other than these sources, there exist many single neutron stars which 
radiate in soft X-ray band with spectra closer to the blackbody compared 
to cooling pulsars. Yet, luminosities of these sources are smaller (total 
L$_x$$\cong$10$^{29}$-10$^{32}$ erg/s), but their number density is very 
large. In spite of the large distances adopted by Guseinov et al. (2003c), 
their birth rate is not smaller than the birth rate of all pulsars and 
about half of the supernova explosion rate. 

For 2 of these sources, RXJ0720.4-3125 and J1308.8+2127, the P values have 
been measured and unconfirmed P values are known for 2 other such sources. 
Note that values of \.{P} were given for the first 2 sources (Zane et al. 
2002; Hambaryan et al. 2002) but these values are uncertain. Since the P 
values of these 2 sources are reliable and their ages are not more than 
10$^6$ yr, it is easy to understand their 
locations on the P-\.{P} diagram for \.{P}$>$10$^{-14}$ s/s. But one faces 
with big difficulties against the high birth rate of such objects. We 
think that dominant number of DITNSs must have P$<$4-5 s. Understanding 
the existence of a relation between some of the DITNSs and AXPs/SGRs 
would be easy if the birth rate of DITNSs with large periods were in 
accordance with the birth rate of AXPs/SGRs, in error limits, which is 
about 60 times smaller than the supernova rate (Guseinov et al. 2003c). 
Since the birth rate of DITNSs is about 30 times larger than the birth 
rate of AXPs/SGRs, only a very small part of DITNSs, which have P$>$6-8 
s, may be related to AXPs/SGRs. 

The known values of temperature and luminosity of DITNSs naturally limit 
the ages of these objects. As the cooling theories which we have used 
in the birth rate estimations do not have large uncertainties for neutron 
stars with masses very close to 1.4 M$_{\odot}$, the ages of most of 
the DITNSs may be close to 10$^6$ yr if their masses are close to 1.4 
M$_{\odot}$. Taking this fact into account, we can begin to construct 
possible evolutionary tracks of DITNSs and AXPs/SGRs. 

The data of the DITNSs may be not so reliable, but even 
in such a case we can not neglect the existence of some neutron stars, for 
example, with P$>$5 s and \.{P}$>$4$\times$10$^{-14}$ s/s for which the 
birth rate may be about 0.1 of the supernova rate. In this case, instead 
of accepting a bimodal number-magnetic field distribution for pulsars, we 
can say that neutron stars with $\tau$$<$10$^4$ yr and B$\sim$10$^{13}$ G 
may transfer to the region of P$>$5 s and \.{P}$>$4$\times$10$^{-14}$ 
s/s mentioned above.

\section{Discussion and Conclusions}
In sections 2 and 3, we have discussed how the lines with constant values 
of $\tau$, B and \.{E} change on the P-\.{P} diagram under the change of 
radius, moment of inertia and the component of the magnetic field 
perpendicular to the rotation axis of pulsar. If pulsars have considerably 
different, but constant, values of R, I and real magnetic field component, 
then their evolution must be similar to what is known. But if these 
parameters change in time, then the values of the braking index at each 
point on the P-\.{P} diagram can change and in such a case the 
evolutionary tracks must be very different and complicated. This 
gives us the possibility to use different parameters of neutron star 
(mass, magnetic field, structure, rotation, propeller mechanism, absence 
of the hydrodynamical equilibrium etc.) to understand the evolution on 
the P-\.{P} diagram and the properties of different types of pulsar.

Today, we face with very different types of neutron star which are located 
on different parts of the P-\.{P} diagram (Figure 1). They may have 
different origins, structures and evolutionary tracks. There are not large 
differences between the locations of radio pulsars and DRQNSs on the 
P-\.{P} diagram and they do not create any significant problem in 
understanding the neutron star characteristics. But in the last about ten 
years AXPs/SGRs and comparably in small degree DITNSs created very 
difficult problems in astrophysics. 

The locations of AXPs/SGRs on the P-\.{P} diagram rise a large 
astonishment, but it is very good that their number density in the Galaxy 
is very small and connected to this their birth rate is also very small. 
This makes it difficult to have necessary amount of observational data but 
also makes it easy to search ways to understand them. On the other hand, 
the $\gamma$-ray burst characteristic makes the problem more difficult. In 
the case of DITNSs, the large difficulties are related to the high birth 
rate of these objects. In this work, we do not discuss the  
difficulties of the magnetar and the accretion from fall back matter 
theories, which are known well. In the magnetar model, the 
existence of the bimodality in the number of pulsars versus the magnetic 
field distribution and the absence of a correlation between the increase 
in \.{P} and the $\gamma$-ray burst activity are significant problems. In 
the accretion from fall back matter theory, the idea of accretion onto 
young neutron stars which may be very active does not seem to be 
realistic and it can not solve any principal problem about AXPs/SGRS 
without taking into account different possible activity of neutron stars 
as mentioned above. Therefore, by rejecting the idea of large magnetic 
field suggested in the magnetar model which is 10$^{14}$-10$^{15}$ G, we 
exclude the bimodality and we accept that different types of neutron star 
have different values of parameter (e.g. different value of mass) as 
represented in the previous sections. 

As known for single radio pulsars, the ratio of the X-ray radiation to the
rate of rotational energy loss lies in the interval
3$\times$10$^{-2}$-10$^{-5}$ (Becker \& Trumper 1997; Possenti et al.    
2002; Becker \& Aschenbach 2002). The name 'anomalous' comes from the
inverse value of L$_x$/\.{E} and DITNS RX J0720.4-3125 is similar to AXPs 
from this point of view (Kaplan et al. 2003; Zane et al. 2002). 
If we consider only L$_x$/\.{E} values, most of the DRQNSs and
DITNSs must belong to inter-class objects between radio pulsars and
AXPs/SGRs. Today, one can begin to investigate the differences and common
properties of single neutron stars in detail. The data show the existence 
of differences not only in the values of the real magnetic
field of neutron stars but also in the masses and the internal structures.
Neutron stars not only have different evolutionary tracks, but they     
may be born with very different values of P and \.{P} also.

There exist models of stable neutron star with masses from 0.1 
M$_{\odot}$ up to 2 M$_{\odot}$ (Bejger \& Haensel 2003), but most of the 
neutron stars have masses of about 1.4 M$_{\odot}$ (Charles \& Coe 2003). 
Actually, this mass interval must be narrower; if the mass is 2 
M$_{\odot}$ the 
rest energy is $\sim$4$\times$10$^{54}$ ergs and the binding energy 
including both the energy carried by neutrinos and the explosion energy 
is $\sim$10\% of the rest energy (Zeldovich \& Novikov 1971). Naturally, 
the energy carried by the neutrino pairs is lower for the collapsed 
matter with smaller mass, but the total energy loss must always be large 
enough (Guseinov 1966) that such energy can not be produced to form a 0.1 
M$_{\odot}$ neutron star for which the binding energy is comparably very 
small. We must also take into account the oscillations which must take 
place under the formation of neutron stars and the masses (the pressure) 
which continue to fall onto their surfaces during the collapse.
Under the collapse of progenitor stars, black holes may also be born that 
the newborn neutron stars must have the supply for stability against the 
overcompression. Neutron stars which have masses very close to the maximum 
mass can not resist against the matter falling on themselves. Therefore, 
the maximum masses of the existing neutron stars must be a little smaller 
than the maximum mass values found from the models and the minimum mass 
must be about 0.5 M$_{\odot}$ which is the value we have adopted for AXP 
1E1841-045 and AXP 1E2259+586 which are connected to SNRs. Fastly rotating 
neutron stars with such small masses may have about 1.3 times larger 
radius compared to the neutron stars with masses close to 1.4 M$_{\odot}$. 

The pulsars with 0.5 M$_{\odot}$, which have the same real magnetic field 
as the massive pulsars with the same spin period, must have about 8 times 
larger value of \.{P} and smaller value of $\tau$ according to the 
expressions (3) and (5). They must also have a bit larger value of 
E$_{el}$ (see expression (7)). As seen from Figure 1, even a magnetic 
field of $\sim$3$\times$10$^{13}$ G is small for such an explanation of 
the positions of all AXPs/SGRs on the P-\.{P} diagram for which the ages 
are not more than $\sim$3$\times$10$^4$ yr. But the small mass model 
gives us several additional possibilities. If the magnetic field axis 
makes an angle close to 90$^o$ with the rotational axis, then the 
atmosphere of the neutron star must be stretched in the direction of the 
magnetic field axis. The extended atmosphere and the high voltage make 
the gas easily flow along the magnetic field lines and the propeller 
mechanism effectively works. From such an approach, it is easy for small 
mass neutron stars with magnetic fields $\ge$10$^{13}$ G to come to the 
positions of AXPs/SGRs on the P-\.{P} diagram in short time. The extended 
low density atmosphere, which is not in stable state, may give many 
possibilities for different activity and also for recombination of the 
magnetic field. It is easier for the particles to be ejected from the 
extended hot atmosphere. In addition to the magnetic field energy, there 
exist rotational and gravitational energies which support the activity of 
AXPs/SGRs. The small mass pulsars must have considerably long cooling time 
and short characteristic time for the rotational energy loss. Therefore, 
the ratio $\frac{L_x}{\dot{E}}$$>$1 can easily be explained by this way. 

In the given approach, it is not difficult to understand DITNSs, because 
they have very simple properties compared to AXPs/SGRs. The increase of 
the rotational period must accelerate the processes of contraction and 
coming to stability. These processes can lead to a considerable increase 
in the braking index. Therefore, many of the DITNSs must have n$>$3. In 
most part of the lifetimes of these sources and AXPs/SGRs, they must 
evolve in the direction of decreasing \.{P}. From this approach, it 
follows that DITNSs must be found in some nearby SNRs. Existence of 
ionized gaseous wind must promote to suppression of the radio pulsar 
phenomenon when their periods are still small. Particularly, large noise 
in the period and possibly the glitch phenomenon must also exist for the 
neutron stars with P$>$6-8 s and \.{P}$>$10$^{-13}$ s/s.

%\final
\clearpage
{\bf Figure Caption} \\
{\bf Figure 1:} Period versus period derivative diagram for different
types of pulsar. The '+' signs denote the radio pulsars with d$\le$3.5
kpc which are connected to SNRs. The 'X' signs show
the positions of the radio pulsars with d$\le$3.5 kpc and
10$^5$$<$$\tau$$<$2$\times$10$^7$ yr which have been detected in X-rays.
The locations of 3 radio pulsars which have d$\le$3.5 kpc and
$\tau$$<$10$^5$ yr are shown with 'circles' to make a comparison between
the birth rates (see text). DITNSs are
represented with 'stars' and DRQNSs are displayed with 'empty squares'.
The 'filled squares' show the positions of all AXPs/SGR in the Galaxy.
Names of DITNSs, DRQNSs, 2 of the AXPs, and
some of the radio pulsars are written. Constant lines of B =
10$^{11-15}$ G, $\tau$ = 10$^{3-9}$ yr, and \.{E} = 10$^{29}$, 10$^{32}$,
10$^{35}$, 3$\times$10$^{35}$ and 10$^{38}$ erg/s are shown. P=10 s line
is also included (see text). This figure is taken from Guseinov et al. 
(2003c).

\clearpage
\begin{figure}[t]
\vspace{3cm}
\includegraphics{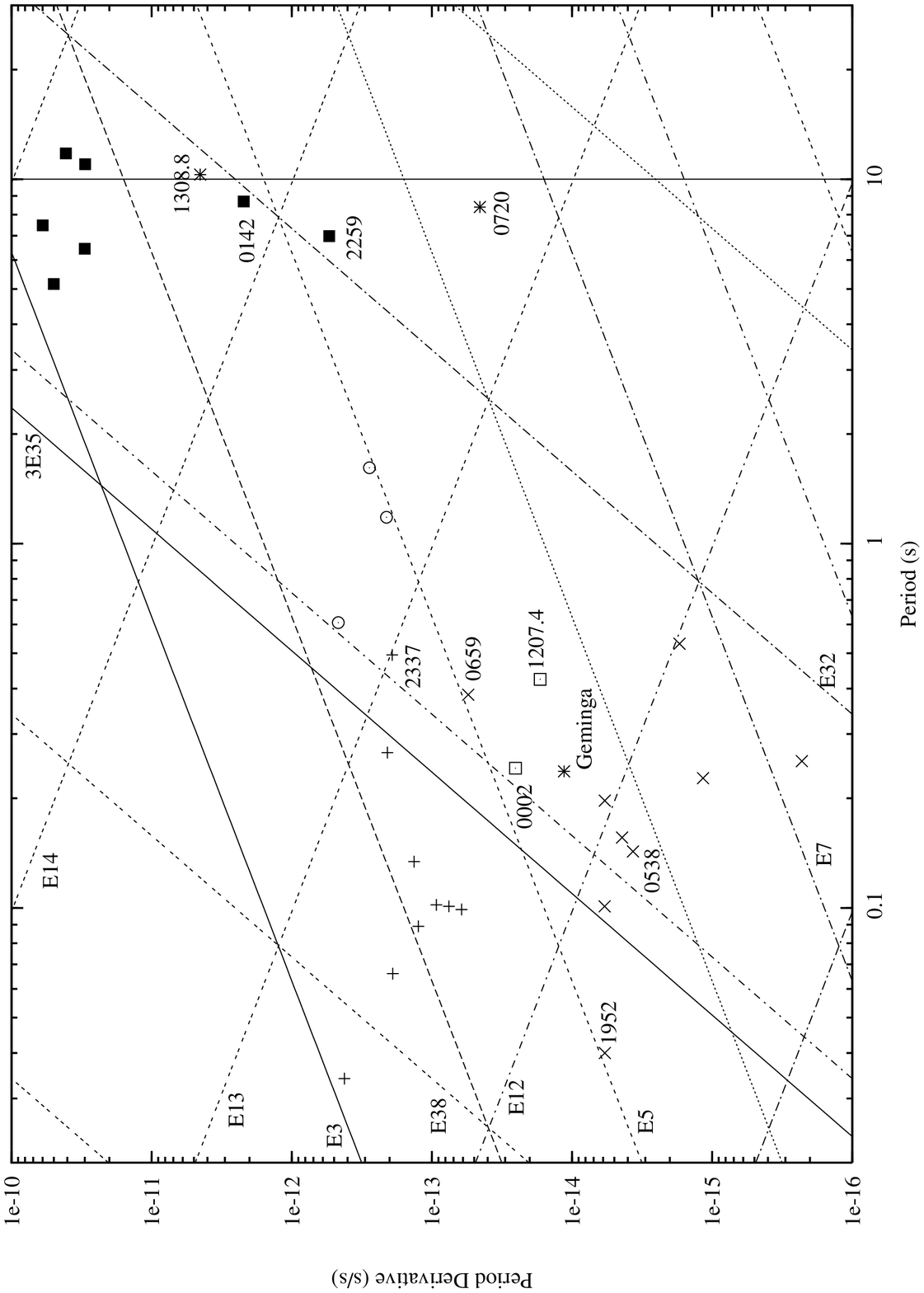}
%\caption*
\end{figure}

\end{document}